\documentclass[preprint,showpacs,floatfix,pre]{revtex4}
\usepackage{amsmath,amssymb}
\usepackage[dvips]{graphicx}

\newcommand{\IFUFG}{Instituto de F{\'{\i}}sica, Universidade Federal de 
Goi{\'a}s, Av. Esperan\c{c}a s/n, 74.690-900, Goi{\^a}nia, GO, Brazil}
\newcommand{\IFMT}{Instituto Federal do Mato Grosso - Campus C{\'a}ceres, 
Av. dos Ramires s/n, 78200-000, C{\'a}ceres, MT, Brazil}
\newcommand{\UFSCAR}{Departamento de F\'{\i}sica, Universidade Federal de S\~{a}o Carlos. 
Rodovia Washington Luiz, km 235, Caixa Postal 676, 13565-905 S\~{a}o Carlos-SP, Brazil}
\newcommand{\UFAM}{Departamento de F\'{\i}sica, Universidade Federal do Amazonas, 3000, Japiim,
69077-000, Manaus-AM, Brazil}
\begin{document}

\author{L. S. Ferreira$^1$, L. N. Jorge$^{1,2}$, A. A. Caparica$^1$, \\
Denise A. do Nascimento$^{3,4}$, J. R. de Sousa$^4$, $^*$Minos A. Neto$^4$\\} 

\address{$^1$ \IFUFG }
\address{$^2$ \IFMT }
\address{$^3$ \UFSCAR }
\address{$^4$ \UFAM}
\email{minos@pq.cnpq.br}

\title{Thermodynamic properties of rod-like chains: entropic sampling 
simulations}

%
%
\date{\today}

\begin{abstract}

In this work we apply entropic sampling simulations to a three-state model which has exact solutions in the microcanonical and 
grand-canonical ensembles. We consider $N$ chains placed on an unidimensional lattice, such that each site may assume one of 
three-states: empty (state 1), with a single molecule energetically null (state 2), and with a single molecule with energy 
$\varepsilon$ (state 3). Each molecule, which we will treat here as dimers, consists of two monomers connected one to each other by a rod.
The thermodynamic properties, such as internal energy, densities of dimers and specific heat were obtained as functions of 
temperature where the analytic results in the micro-canonical and grand-canonical ensembles were successfully confirmed by 
the entropic sampling simulations.

\textbf{Keywords}: rod-like chains, entropic sampling simulations, Wang-Landau algorithm

\end{abstract}

\maketitle

\section{Introduction\protect\nolinebreak}

In recent years unidimensional models of polymers have received a special 
attention. By treating a simple version of reality, these models can give us 
extremely relevant information on more complex problems. One of such examples 
is the problem of polydisperse chains of polymers on a unidimensional
lattice \cite{minos2006}. This kind of model is a good example to explain 
equilibrium polymerization \cite{wheeler1980,wheeler1981}, living polymers 
\cite{dudowicz1999} and in the phase transition of liquid sulfur 
\cite{greer1998}. More recently this problem has been treated in Bethe 
\cite{minos2008} and Husimi \cite{minos2013} lattices. 

Water is a special fluid of great biological relevance with many 
technological applications. Recently, da Silva \textit{et al.} 
\cite{dasilva2015} analyzed the thermodynamics and kinetic unidimensional 
lattice gas model with repulsive interaction. In this model the residual entropy 
and water-like anomalies in density are investigated using matrix technique and 
Monte Carlo simulations.

Another interesting problem is the unidimensional model of a solvent with $q$ 
orientational states to explain the effects of hydrophobic interaction 
\cite{kolomeisky1999}. Initially, the model was described by Ben-Naim 
\cite{ben-naim} for a unidimensional model with many states (related to 
the $q$ state unidimensional Potts model \cite{potts}), which can be adapted 
to illustrate the entropy of dimer chains placed on an unidimensional 
lattice with $q$ states \cite{denise2015}. Another example of this type are 
molecules with multiple adsorption states \cite{quiroga2013}. In such models 
the thermodynamic functions are used to describe the adsorption of antifreeze 
proteins onto an ice crystal.

In such a scenario, models presenting three-states have many applications. 
For example, the Ising model has been used for a long time as a ``toy model'' 
for diverse objectives, as to test and to improve new algorithms and methods of 
high precision for the calculation of critical exponents in equilibrium 
statistical mechanics using the Monte Carlo methods as Metropolis 
\cite{metropolis1953}, Swendsen-Wang \cite{swendsen1987}, single histogram \cite{ferrenberg1988}, broad
histogram \cite{oliveira1996}, and Wang-Landau \cite{wang2001} methods. This model has been applied in 
various fields of science as for example nucleation on complex networks (statistical mechanics) \cite{chen2015}, 
religious affiliation (social systems) \cite{mccartey2015}, kinetic model to 
analyze tax evasion dynamics (econophysics) \cite{nuno2014}, drugs as 
benzodiazepine (biochemistry) \cite{polc1982}, deterministic epidemic model 
(epidemiology) \cite{capasso1978}, dynamical system for cancer virotherapy 
\cite{dingli2006} (medicine) and kinetic models of molecular association 
(molecular and cellular biophysics) \cite{jackson2006}. 

We will consider in this work a three-state unidimensional system with 
empty sites, energetically null single molecules, and single molecules with 
energy $\varepsilon$, where it is important to emphasize that the molecules are
non-interacting.  In Section II we present the model and formalisms. The entropic 
sampling simulations are described in Section III. The results and discussions are 
presented in Section IV. Finally, the last section is devoted to ultimate remarks and conclusions.

\section{Models and Formalism}\label{models}

\subsection{Model}
We consider an unidimensional model where each molecule consists of two monomers
linked by a rod, which we call dimer. The number of dimers may vary as $0\leq N\leq L$, where 
$L$ is the lattice size. Each site may assume one of three states, 
namely, an empty site, a molecule energetically null, or a molecule with 
energy $\varepsilon$. In each case, the energy of the state is therefore 0, 
0, and $\varepsilon$, respectively. In addition, in this model there is no 
interaction between dimers, nor between the monomers that form the dimers.

The Hamiltonian is written as
\begin{equation}
 \mathcal{H}=\varepsilon\sum_{i=1}^{L} \delta_{3,q_i} ,
 \label{hamiltoniana1}
\end{equation}
where the sum extends over all sites of the lattice and $q_i$ represents the state
of the site.

The ground state consists in the lattice being completely empty or with any number
of dimers energetically null, the next energy level corresponds to a single dimer with energy 
$\varepsilon$ and any number of dimers energetically null and so on, until the configuration with 
maximum energy where each site is occupied by a dimer with energy $\varepsilon$. 

\begin{figure}[h]
 \begin{center}
 \includegraphics[angle=0,scale=0.5]{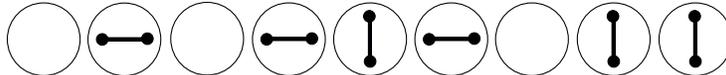}
\end{center}
 \caption{Scheme of a lattice with $L=9$ sites for a possible configuration. We have here three empty sites, three 
 sites occupied by dimers energetically null and three by dimers with energy $\varepsilon$.}
 \label{fig.dimer}
\end{figure}

In Fig.\ref{fig.dimer} we show a possible configuration for $L=9$ and $N=6$. Each circle with a 
horizontal bar represents a dimer energetically null, the circles with vertical 
bars correspond to dimers with energy 
$\varepsilon$ and the open circles are for empty sites.
 
\subsection{Microcanonical solution}

The microcanonical formulation of the model of dimer chains on an 
unidimensional lattice with two types of dimers is done as follows. 
Let $N_1$, $N_2$, and $N_3$ be the fixed numbers of empty sites, dimers with zero energy, 
and dimers with energy $\varepsilon$, respectively. According to 
\eqref{hamiltoniana1}, the energy of the system can be written as
\begin{equation}
 \mathcal{H}=N_3\varepsilon , 
 \label{hamiltoniana2}
\end{equation}

The number of ways one can allocate $N_1$ empty sites, $N_2$ dimers of type 1, 
and $N_3$ dimers of type 2 on an unidimensional lattice of size $L$ is 
given by 
\begin{equation}
 \Omega=\frac{L!}{N_1!N_2!N_3!}.
 \label{omega1}
\end{equation}
Let $N=N_2+N_3$ be the total number of dimers and $U=N_3\varepsilon$ the 
total energy. We can therefore rewrite the number of configurations 
\eqref{omega1} as a function of $U,$ $N,$ and $L$ as
 \begin{equation}
 \Omega(U,N,L)=\frac{L!}{(L-N)!(N-U/\varepsilon)!(U/\varepsilon)!}.
 \label{omega2}
\end{equation}
Now, calculating the entropy via logarithms and the Stirling's 
formula, we obtain
\begin{equation}
 s(u, \rho)=-k_B(1-\rho)\ln (1-\rho)-k_B(\rho-u/\varepsilon)\ln(\rho-u/\varepsilon)-k_B(u/\varepsilon)\ln(u/\varepsilon) ,
\label{entropy}
\end{equation}
where $s=\frac{S}{L}$ is the entropy per site, $k_B$ is the Boltzmann's constant, $u =\frac{U}{L}$ is the 
energy per site, and $\rho=\frac{N}{L}$ is the density of dimers. 

The equations of state in the entropy representation are $\frac{1}{T}=\left(\frac{\partial s}{\partial u}\right)_{\rho}$ and
$-\frac{\mu}{T}=\left(\frac{\partial s}{\partial \rho}\right)_{u}$, where $T$ is the absolute temperature and $\mu$ is the
chemical potential. Solving these two equations we obtain the energy 
\begin{equation}
 u=\frac{\varepsilon e^{(\mu-\varepsilon)/k_BT}}{1+e^{\mu/k_BT}+e^{(\mu-\varepsilon)/k_BT}}.
 \label{e_micro}
\end{equation}
and the density of dimers
\begin{equation}
 \rho=\frac{e^{\mu/k_BT}+e^{(\mu-\varepsilon)/k_BT}}{1+e^{\mu/k_BT}+e^{(\mu-\varepsilon)/k_BT}},
 \label{rho1}
\end{equation}
as functions of $T$.

\subsection{Transfer matrix technique: grand canonical ensemble}

In the grand canonical ensemble the grand partition function is given by
\begin{equation}
\Xi = \sum_{j} \exp (-\beta \mathcal{H}_{j}+\beta \mu N_{j}).
\end{equation}
where $\beta=1/k_BT$, $\mu$ is the chemical potential of the dimer, and the sum is over all possible configurations.

Since each state of the system is characterized by a set of lattice variables $q_i$, 
which may assume one of three states, the grand partition function for our model
can be written as

\begin{equation}\label{xin3n}
\Xi= \sum_{\{q_i\}}\exp [-\beta \varepsilon N_3 +\beta \mu N],
\end{equation}
or
\begin{equation}
\Xi= \sum_{\{q_i\}}\exp \left[-\frac{\beta \varepsilon}{2} \sum_{i=1}^{L} (\delta_{3,q_i}+\delta_{3,q_{i+1}})+
\frac{\beta \mu}{2} \sum_{i=1}^{L} (\delta_{2,q_i}+\delta_{2,q_{i+1}}+\delta_{3,q_i}+\delta_{3,q_{i+1}})\right],
\end{equation}
where $\{q_i\}$ denotes that the sum runs over all possible configurations and, using periodic boundary conditions, 
we adopted a double sum over all sites divided by $2$ in order to 
obtain the usual notation of the transfer matrix technique. More concisely we have
\begin{equation}
\Xi=\sum_{\{q_i\}} \prod_{i=1}^{L} T_{q_i~q_{i+1}},
\end{equation}
where
\begin{equation}\label{tqq}
T_{q_i~q_{i+1}}=\exp \left[-\frac{\beta}{2}(\varepsilon-\mu)(\delta_{3,q_i}+\delta_{3,q_{i+1}})+
\frac{\beta \mu}{2}(\delta_{2,q_i}+\delta_{2,q_{i+1}})\right].
\end{equation}

The transfer matrix is therefore defined by \eqref{tqq} and can be written as
\[ T =
\left(
\begin{array}{ccc}
T_{11} & T_{12} & T_{13} \\
T_{21} & T_{22} & T_{23} \\
T_{31} & T_{32} & T_{33} 
\end{array} 
\right)=
\left(
\begin{array}{ccc}
1 & z^{1/2} & (wz)^{1/2} \\
z^{1/2} & z & w^{1/2}z \\
(wz)^{1/2} & w^{1/2}z & wz 
\end{array} 
\right),\]
where $z=e^{\mu\beta}$ and $w=e^{-\varepsilon\beta}$. The largest eigenvalue of this matrix is
\begin{equation}
 \lambda=1+z+wz,
 \label{lambda}
\end{equation}
which corresponds to the one site grand canonical partition function: $\Xi=\lambda^L$.

\subsection{Thermodynamic quantities}\label{quantities}

The grand canonical potential per site is given by
\begin{equation}
 \phi=-k_BT\ln\lambda. \label{eq.phi}
\end{equation}
The entropy follows from the equation of state $s=-\frac{\partial \phi}{\partial T}$, and using
$\frac{\mu}{k_BT}=\ln z$ and $\frac{\mu-\varepsilon}{k_BT}=\ln wz$, we obtain an expression for the entropy
\begin{equation}
 s=k_B\ln \lambda -k_B\frac{z(1+w)}{\lambda}\ln z -k_B\frac {wz}{\lambda}\ln w.
\end{equation}
Assuming that the largest eigenvalue of $\lambda$ of the transfer matrix is not degenerate, the density of dimers as function of 
$\lambda$ is written as
\begin{equation}
\rho=\frac{z}{\lambda}\frac{\partial \lambda}{\partial z}=\frac{z(1+\omega)}{\lambda}, 
\label{rho}
\end{equation}
and using the definition of mean energy in the grand canonical ensemble
\begin{equation}
u=-\frac{\partial \ln\lambda}{\partial \beta}+\frac{\mu}{\beta}\frac{\partial \ln\lambda}{\partial \mu}=\varepsilon\frac{wz}{\lambda}, 
\label{u_gran}
\end{equation}
we can write $z$ and $w$ as functions of $\rho$ and $u$, respectively and 
obtain an expression to the entropy identical to \eqref{entropy}.
By replacing $z$ and $\lambda$ in \eqref{rho} and \eqref{u_gran} we obtain the same expressions
of \eqref{rho1} and \eqref{e_micro}, respectively, thus confirming the equivalence between the
microcanonical and the grand canonical formalisms.

From \eqref{xin3n} it follows that $\bar{N_3}= -\frac{L}{\beta}\frac{\partial\ln\lambda}{\partial\varepsilon}$, giving
\begin{equation}
 n_3=\frac{\bar{N_3}}{L} =\frac{wz}{\lambda}=
 \frac{e^{(\mu-\varepsilon)/k_BT}}{1+e^{\mu/k_BT}+e^{(\mu-\varepsilon)/k_BT}},
\end{equation}
and using \eqref{rho} we have
\begin{equation}
n_2=\rho-n_3=\frac{z}{\lambda}=\frac{e^{\mu/k_BT}}{1+e^{\mu/k_BT}+e^{(\mu-\varepsilon)/k_BT}},
\end{equation}
and 
\begin{equation}
n_1=1-\rho=\frac{1}{\lambda}=\frac{1}{1+e^{\mu/k_BT}+e^{(\mu-\varepsilon)/k_BT}}.
\end{equation}

The specific heat can be obtained from the definition $c=\frac{du}{dT}-\mu\frac{d\rho}{dT}$, giving
\begin{equation}\label{eq.cv}
 c(T,\varepsilon,\mu)= \frac{e^{\mu/k_BT}}{k_BT^2}\dfrac{(\varepsilon-\mu)^2e^{-\varepsilon/k_BT} 
 +\varepsilon^2 e^{(\mu-\varepsilon)/k_BT}+\mu^2}{(1+e^{\mu/k_BT}+e^{(\mu-\varepsilon)/k_BT})^2}. 
\end{equation}
with an explicit dependence on $T$, $\varepsilon$ and $\mu$. It is noteworthy that \eqref{eq.cv} is 
invariant with respect to permutation of the variables $\varepsilon$ and $\mu$. 

\section{Simulations}\label{simulations}

We apply to our model an entropic sampling simulation based on Wang-Landau 
sampling \cite{wang2001}, taking into account the improvements prescribed in 
\cite{caparica2012,caparica2014}. Our model has two degrees of freedom: the 
energy $U$ and the total number of dimers $N$. Accordingly we should seek 
a joint density of states $g(U,N)$. We define an unidimensional lattice 
of length $L$ and each site can be in any of three states: 1 - empty, 2 - occupied by 
an energetically null dimer, and 3 - filled by a dimer with energy 
$\varepsilon$. A trial move is defined as giving sequentially 
to each site the possibility of changing with identical probability to any new 
state, including remaining in the same state. A Monte Carlo sweep is defined as 
a sequence of $L$ trial moves. At the beginning of the simulation we set 
$g(U,N)=1$ for all energy levels and numbers of particles. The random walk runs 
through all the energy levels and numbers of particles with a probability
\begin{equation}
p(U \rightarrow U', N \rightarrow 
N')=\text{min}\left\{\frac{g(U,N)}{g(U',N')},1\right\}
\end{equation}
where $U,N$ and $U',N'$ are the energies and numbers of particles of the current 
and the new attempting configuration. The density of states and the histogram 
are updated after each Monte Carlo step and the histogram is considered flat if 
$H(U,N)>0.8\left\langle H\right\rangle$ for all energies and numbers of 
particles, where $\left\langle H\right\rangle$ is an average over energies and 
numbers of particles. In order to estimate the mean values of the number of 
dimers of each type and of the empty sites, we accumulate microcanonical 
averages of these quantities, but these sums begin only from the 7th 
Wang-Landau level onwards \cite{caparica2012}. The simulations were halted 
using the checking parameter that verifies the convergence of the peak of the 
heat capacity to a steady value \cite{caparica2014}. The grand-canonical 
averages of any thermodynamic variable $A$ can be calculated as

\begin{equation}\label{mean}
 \bar{A}(T,\mu)=\dfrac{\sum_{U,N}\langle A\rangle_{U,N} g(U,N) 
  e^{-\beta(U-\mu N)}}{\sum_{U,N} g(U,N) e^{-\beta(U-\mu N)}} ,
\end{equation}
where $\langle A\rangle_{U,N}$ is the microcanonical average accumulated during 
the simulations.

\section{Results and Discussion}\label{results}

\begin{figure}[!ht]
 \begin{center}
 \includegraphics[angle=-90,scale=0.5]{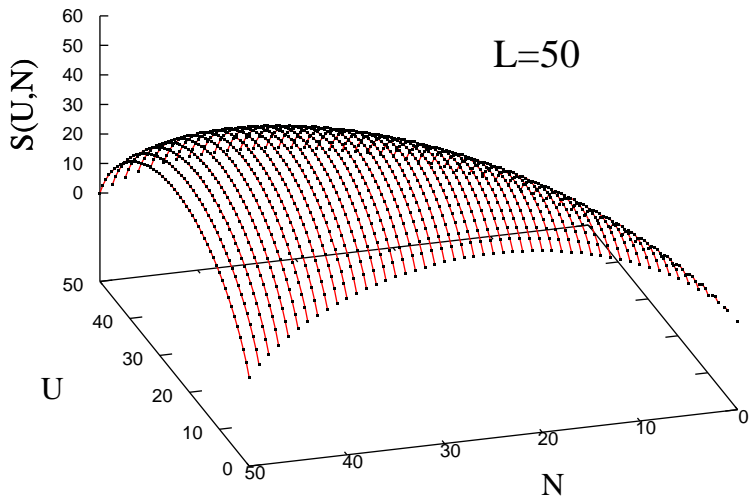}
\end{center}
 \caption{(Color online)Logarithm of the density of states for $L=50$ calculated analytically (lines) and obtained by entropic
 simulations (dots) for ten independent runs.}
 \label{fig.logdos}
\end{figure}

As described in Section \ref{models}, the thermodynamic properties of our model are size independent. In this
study we chose the lattice size $L=50$. To obtain analytic results for the entropy $S(U,N)$, we take the 
logarithm of \eqref{omega2}, with the energy varying from $U=0$ to $U=N$, where we set $\varepsilon=1$, and the number of 
dimers varying from $N=0$ to $N=50$. 
The entropic simulations were performed following the prescriptions of Section \ref{simulations}. 
In Fig. \ref{fig.logdos} we present a comparison between the two ways of obtaining the density of states, 
where the continuous lines represent the analytical solution and the dots are the average
over ten independent runs. We can see that the lines and the dots coincide within very small error bars. 
Once obtained the density of states we can calculate any thermodynamic average using \eqref{mean}. 

\begin{figure}[!ht]
 \begin{center}
 \includegraphics[angle=-90,scale=0.5]{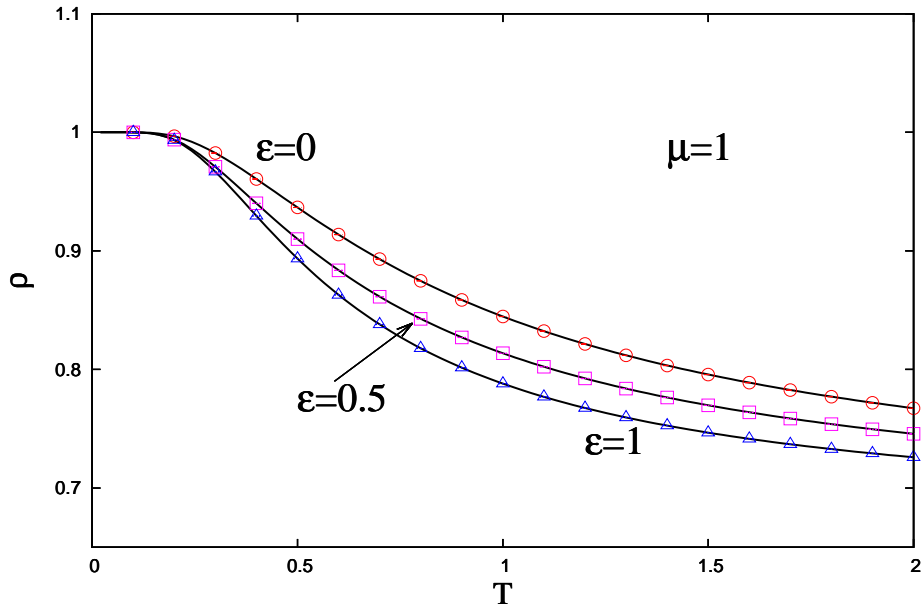}
\end{center}
\caption{(Color online) Dependence of the density of dimers on the temperature for
 $\mu=1$ and $\varepsilon=0, 0.5, 1$. The lines represent the exact results and the dots the simulations.}
 \label{n2}
\end{figure}

In Fig. \ref{n2} we present the dependence of the density of dimers on the temperature for $\mu=1$ and $\varepsilon=0, 0.5, 1$. At $T=0$ all sites are occupied 
by dimers and the density of dimers decreases with increasing temperature and this decrease is more pronounced for larger $\varepsilon$.

\begin{figure}[!ht]
 \begin{center}
 \includegraphics[angle=-90,scale=0.5]{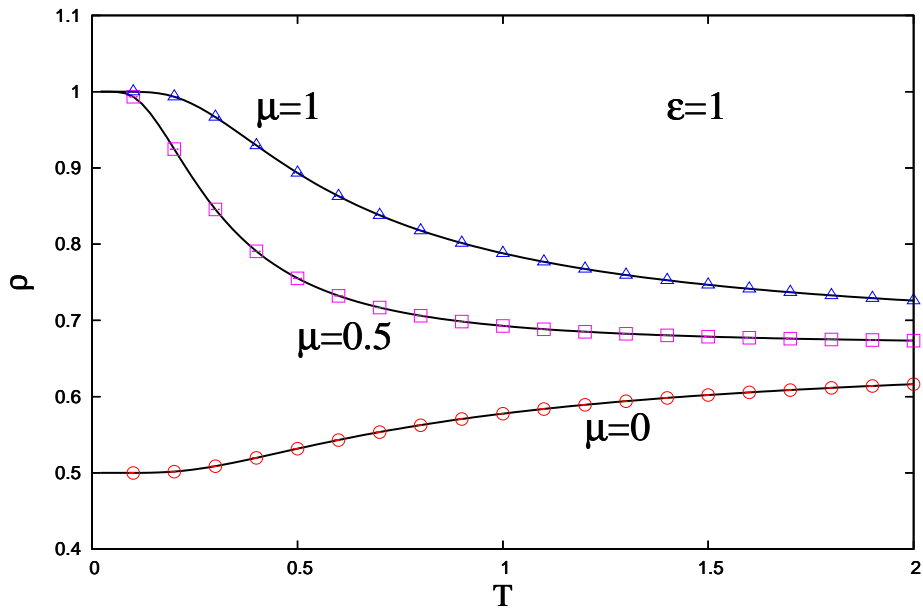}
\end{center}
\caption{(Color online) Dependence of the density of dimers on the temperature for
 $\varepsilon=1$ and $\mu=0, 0.5, 1$. The lines represent the exact results and the dots the simulations.}
\label{n3}
\end{figure}

In Fig. \ref{n3} we show the temperature dependence of the density of dimers for $\varepsilon=1$ and $\mu=0, 0.5, 1$. 
At zero temperature all sites are occupied if $\mu\neq0$, but for $\mu=0$ half lattice is empty. The density of
dimers decreases with temperature for $\mu\neq0$ and this decrease is more pronounced for smaller $\mu$. Nevertheless
for $\mu=0$ the density of dimers increases with temperature. In all situations $\rho\rightarrow2/3$ as $T\rightarrow\infty$.

\begin{figure}[!ht]
 \begin{center}
 \includegraphics[angle=-90,scale=0.5]{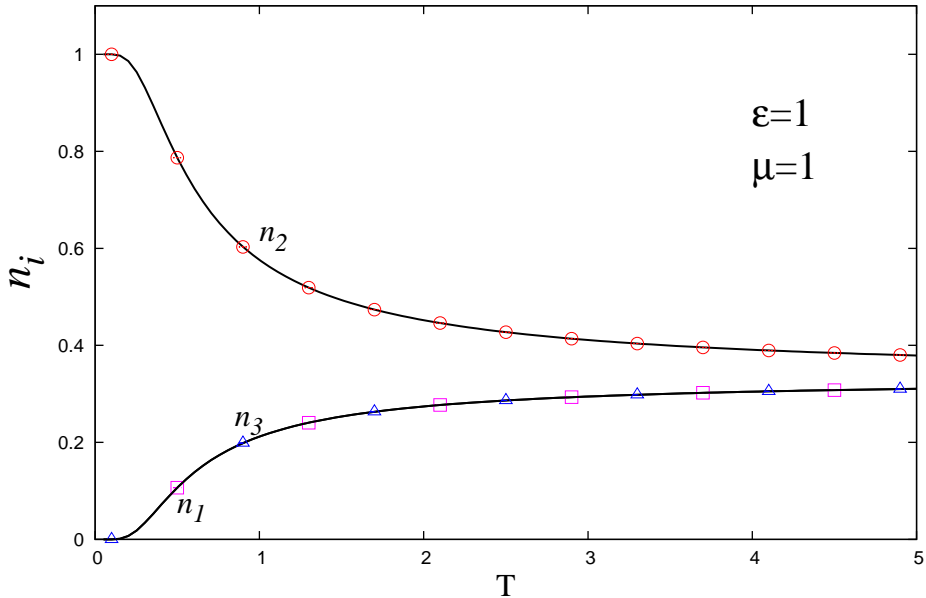}
\end{center}
\caption{(Color online) Temperature dependence of the densities of dimers $n_2$ and $n_3$, and the 
density of empty sites $n_1$. The lines represent the exact results and the dots the simulations.}
 \label{n123}
\end{figure}

Fig. \ref{n123} shows the densities of empty sites $n_1$, energetically null dimers $n_2$ and energetic dimers 
$n_3$ when $\mu=\varepsilon$. At zero temperature all sites are occupied by energetically null dimers. The 
number of empty sites and energetic dimers increases equally with increasing temperature while $n_2$ decreases.

\begin{figure}[!h]
 \begin{center}
 \includegraphics[angle=-90,scale=0.5]{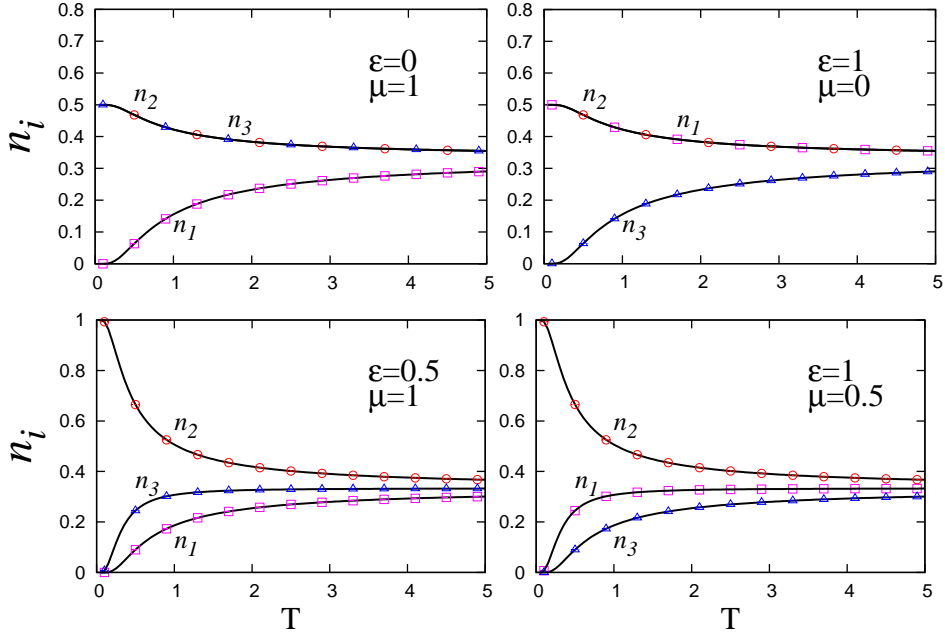}
\end{center}
\caption{(Color online) Dependence on temperature of the densities $n_1$, $n_2$ and $n_3$ for different values of $\varepsilon$ and $\mu$.
 The lines represent the exact results and the dots the simulations.}
 \label{n4}
\end{figure}

In Fig. \ref{n4} we show the behavior of the densities $n_1$, $n_2$ and $n_3$ for different values
of $\varepsilon$ and $\mu$. If $\varepsilon=0$, at $T=0$ the lattice is equally fulfilled by both types of
dimers, since in fact in this case they are energetically equivalent. When $\mu=0$, at zero temperature the 
lattice is half empty and the other half is occupied by dimers energetically null. With increasing temperature
the density of energetic dimers increases and the number of empty sites and dimers energetically null decreases
equally. When both $\mu$ and $\varepsilon$ are not null, at $T=0$ the lattice is completely fulfilled by dimers
energetically null. The increase of temperature favors the increase of the density of energetic dimers if 
$\varepsilon<\mu$ or empty sites, if $\varepsilon>\mu$. All three densities $n_1$, $n_2$ and $n_3$ tend to 1/3 
when the temperature tends to infinite, so that at high temperatures the lattice is equally occupied by energetic dimers, 
energetically null dimers and vacancies.

\begin{figure}[!h]
 \begin{center}
 \includegraphics[angle=-90,scale=0.5]{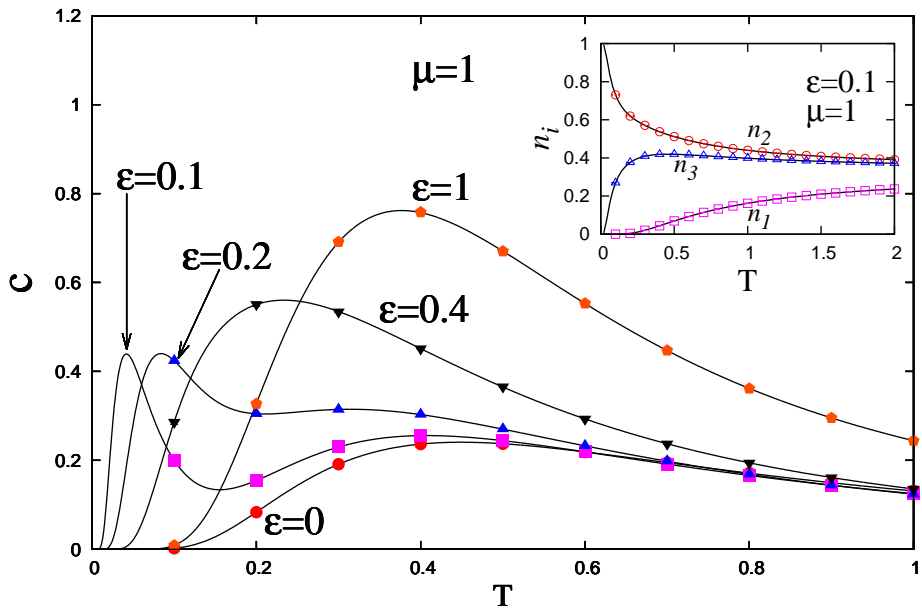}
\end{center}
\caption{(Color online) Specific heat per site for $\mu=1$ and $\varepsilon=0, 0.1, 0.2, 0.4,$ and $1$. The continuous lines represent the exact 
analytic results and the symbols are the mean over ten independent runs for $L=50$. In the inset we show the dependence of $n_1$,
$n_2$, and $n_3$ with temperature for $\mu=1$ and $\varepsilon=0.1$ where it is evident the sudden increase of the density of energetic
dimers at low temperatures.}
\label{cv}
\end{figure}

Finally in Fig. \ref{cv} we depict plots of the specific heat with $\mu=1$ and $\varepsilon=0, 0.1, 0.2, 0.4$ and $1$.
The continuous lines represent the exact analytic results given by \eqref{eq.cv}, and the symbols are the mean over ten 
entropic sampling results for the specific heat expressed in terms of energy fluctuations 
\begin{equation}
 c=\frac{1}{LT^2}(\langle E^2 \rangle - \langle E \rangle^2), 
\end{equation}
where $E=U + \mu N$. The agreement with the exact results are excellent and the error bars are less than the symbols.
An interesting effect one can observe is the emergence of a peak of the specific heat at low temperatures when $\varepsilon<<\mu$.
This collective behavior is due to the low energetic cost of the particles entering the system. In bottom-left of Fig. \ref{n4} we
see the increase of $n_3$ with increasing temperature. In the inset of Fig. \ref{cv} we show the plots of $n_i$ for $\mu=1$ and
$\varepsilon=0.1$. The sudden increase of $n_3$ is evident. At the end of subsection \ref{quantities} we pointed out that \eqref{eq.cv}
is invariant under permutation of $\varepsilon$ and $\mu$. As a result, if we set $\varepsilon=1$ and plot the specific heat for 
$\mu=0.1, 0.2, 0.4,$ and $1$ we obtain an identical graph with the first sharp peak occurring for $\mu=0.1$ at low temperatures.
In this case, as can be observed at bottom-right of Fig. \ref{n4} we have a pronounced increase of vacancies (decrease of the dimer
density) at low temperatures for $\mu<<\varepsilon$. Since $c=\frac{du}{dT}-\mu\frac{d \rho}{dT}$, we see that the first term is dominant when $\varepsilon<<\mu$, while the second
prevails when $\mu<<\varepsilon$.

\section{Conclusions}\label{conclusions}

Unidimensional models of polymers as the problem of polydisperse chains of polymers \cite{minos2006}, kinetic 
unidimensional lattice gas model with repulsive interaction \cite{dasilva2015} and the model of a solvent with $q$ 
orientational states to explain the effects of hydrophobic interaction \cite{kolomeisky1999} are some examples 
of statistical mechanics models that can illustrate some real problems applications. 

We carried out entropic sampling simulations of a simple unidimensional model of molecules that are 
constituted of a rod and two monomers (dimers), which has exact solutions in both the microcanonical and the 
grand-canonical ensembles, where we have shown the equivalence between ensembles.  

We have obtained quite accurate simulational results as compared with the available analytical exact expressions for the 
thermodynamic properties such as entropy, densities of dimers, and specific heat. The specific heat exhibits the typical 
behavior of a tail proportional to $1/T^{2}$ in the high temperatures limit. Another important point is that as $T\rightarrow0$ the 
specific heat also tends to zero, not violating the third law of thermodynamics. The specific heat as a function of temperature 
presents a second rounded maximum, when $\varepsilon<<\mu$ or a unique rounded maximum, when $\varepsilon\sim\mu$. This effect 
is known as a ``Schottky hump'' that when observed in experimental situations is a hint that there are two privileged states 
in the system as is the case in our model. 

Finally, the entropic sampling simulation applied to our model proved to be very efficient when compared with the analytical 
results indicating that it can be adopted in the future to more complex systems such as, for instance, a generalization of 
the model of a solvent with $q$ orientational states to explain the effects of hydrophobic interaction \cite{kolomeisky1999} 
and the thermodynamics and kinetic gas model with repulsive interaction \cite{dasilva2015}, in bi- and tridimensional systems.

\vspace{1.0cm}
\textbf{ACKNOWLEDGEMENT}

We acknowledge the computer resources provided by LCC-UFG and IF-UFMT. L. N. 
Jorge acknowledges the support by FAPEG, L. S. Ferreira the support by CAPES and Minos A. Neto the support by CNPq.


\begin{thebibliography}{14}

\bibitem{minos2006} J. F. Stilck \textit{et al.}, \textit{Physica A} \textbf{368} 442 (2006).

\bibitem{wheeler1980} J. C. Wheeler \textit{et al.}, \textit{Phys. Rev. Lett.} \textbf{45} 1748 (1980).
 
\bibitem{wheeler1981} J. C. Wheeler and S. J. Pfeuty, \textit{Phys. Rev. A} \textbf{24} 1050 (1981).

\bibitem{dudowicz1999} J. Dudowicz \textit{et al.}, \textit{J. Chem. Phys.} \textbf{111} 7116 (1999).

\bibitem{greer1998} S. C. Greer, \textit{J. Phys. Chem. B} \textbf{102} 5413 (1998).

\bibitem{minos2008} Minos A. Neto and J. F. Stilck, \textit{J. Chem. Phys.} \textbf{128}, 184904 (2008).

\bibitem{minos2013} Minos A. Neto and J. F. Stilck, \textit{J. Chem. Phys.} \textbf{138}, 044902 (2013).

\bibitem{dasilva2015} Fernando Barbosa V. da Silva \textit{et al.}, \textit{J. Chem. Phys.} \textbf{142}, 144506 (2015).

\bibitem{kolomeisky1999} A. B. Kolomeisky and B. Widom, \textit{Faraday Discuss.} \textbf{112}, 81 (1999).

\bibitem{ben-naim} A. Ben-Naim. \textit{Thermodynamics for Chemists and Biochemists}. (Plenum, New York, 1992).

\bibitem{potts} F. Y. Wu, \textit{J. Appl. Phys.} \textbf{55} 2421 (1984).

\bibitem{denise2015} Denise A. do Nascimento \textit{et al.}, \textit{Physica A} \textbf{424} 19 (2015).

\bibitem{quiroga2013} E. Quiroga and A. J. Ramirez-Pastor, \emph{Chem. Phys. Lett.} \textbf{556}, 330 (2013).
 
\bibitem{metropolis1953} N. Metropolis \textit{et al.}, \textit{J. Chem. Phys.} \textbf{21} 1087 (1953).

\bibitem{swendsen1987} R. H. Swendsen and J. -S. Wang, \textit{Phys. Rev. Lett.} \textbf{58} 86 (1987).
 
\bibitem{ferrenberg1988} A. M. Ferrenberg and R. H. Swendsen, \textit{Phys. Rev. Lett.} \textbf{61} 2635 (1988).

\bibitem{oliveira1996} P. M. C. de Oliveira \textit{et al.}, \textit{Braz. J. Phys.} \textbf{26} 677 (1996).

\bibitem{wang2001} F. Wang and D. P. Landau, \textit{Phys. Rev. Lett.} \textbf{86} 2050 (2001).
 
\bibitem{chen2015} H. Chen and C. Shen, \textit{Physica A} \textbf{424} 97 (2015). 

\bibitem{mccartey2015} M. Mccartney and D. H. Glass, \textit{Physica A} \textbf{419} 145 (2015).

\bibitem{nuno2014} N. Crokidakis, \textit{Physica A} \textbf{414} 321 (2014).

\bibitem{polc1982} P. Polc \textit{et al.}, \textit{Naunyn-Schmiedeberg's Arch. Pharmacol.} \textbf{321} 260 (1986).

\bibitem{capasso1978} V. Capasso and G. Serio, \textit{Math. Biosci.} \textbf{42} 43 (1978).

\bibitem{dingli2006} D. Dingli \textit{et al.}, \textit{Math. Biosci.} \textbf{199} 5 (2006).

\bibitem{jackson2006} Meyer B. Jackson \textit{Molecular and Cellular 
Biophysics}. (Cambridge University Press, 2006).

\bibitem{caparica2012} A.A. Caparica and A.G. Cunha-Netto, Phys. Rev. E 
\textbf{85}, 046702 (2012).

\bibitem{caparica2014} A.A. Caparica, Phys. Rev. E \textbf{89}, 043301 (2014).

\end{thebibliography}
\end{document}